\documentclass[a4paper,11pt]{article}
\usepackage{pos}
\usepackage{graphicx}
\usepackage{bm}

\title{Towards understanding of the inclusive vs exclusive puzzle in
  the $|V_{cb}|$ determinations}  
\author*[a,b]{Shoji Hashimoto}
\affiliation[a]{Institute for Particle and Nuclear Studies,
  High Energy Accelerator Research Organization (KEK),
  Tsukuba, Ibaraki 305-0801, Japan}

\affiliation[b]{SOKENDAI (The Graduate University for Advanced Studies),
  Tsukuba, Ibaraki 305-0801, Japan}

\emailAdd{shoji.hashimoto@kek.jp}
\abstract{
  Lattice calculation may play an important role to understand the
  cause of the long-standing puzzle among the determination of
  $|V_{cb}|$.
  The key element is the computation of the inclusive decay rate,
  for which the formalism is under development.
  The method has its own problem: the approximation of the phase-space
  factor and finite volume effect.
  I discuss the current status of such study and future prospects.
}

\FullConference{European network for Particle physics, Lattice field theory and Extreme computing (EuroPLEx2023)\\
 11-15 September 2023\\
Berlin, Germany\\}

\begin{document}
\maketitle

\section{The puzzle}
It has been more than a decade since the tension between the exclusive
and inclusive determinations of $|V_{cb}|$ and $|V_{ub}|$ is
recognized
(see \cite{Dingfelder:2016twb} for a review).
Other than various experimental issues, the exclusive method relies on
the lattice QCD calculation of the relevant form factors such as
$B\to D^{(*)}\ell\bar\nu$ or $B\to\pi\ell\bar\nu$, for which some
potential problems are discussed in a companion talk
\cite{Hashimoto_semileptonic},
while the inclusive determination uses the perturbative QCD method
supplemented by the operator product expansion (OPE) to calculate the
semi-leptonic decay rate at the quark level, which corresponds to the
sum over all possible final states.
In OPE, one needs matrix elements of the operators sandwiched by the
initial $B$ meson states; they can be determined in principle by
fitting the experimental data of the integrated decay rate with
various kinematical cuts.
Theoretically, both the exclusive and inclusive methods are considered
reliable and systematically improvable, so that the tension poses a
tremendous challenge for the theorists (assuming that the systematic
uncertainties in the experimental analysis is reliable).

Ideally, the puzzle can be solved if the decay rate was obtained both 
experimentally and theoretically in the entire phase space of the
semi-leptonic decays, {\it i.e.} 
the invariant mass $m_X$ of the final hadronic state and its 
recoil momentum $\bm{q}^2$ (apart from the angular distributions),
and if the determinations of $|V_{xb}|$ at each point of the phase
space all agreed.
But, theoretically the calculation of the differential decay rate is a
formidable task even with the first-principles simulation of lattice
QCD, especially for excited hadronic final states.
More tractable is a non-perturbative lattice calculation of the
integrated decay rate over some phase space, and it is the subject of
this talk.
The integral over the phase space reduces to the inclusive rate when
the entire phase space is covered.
There are many other observables such as those with kinematical cuts
or some moments of $m_X$ and so on.

In the following sections, I briefly outline the formalism to compute
such integrated rates in lattice QCD.
Then, some discussions of potential systematic errors and future
prospects follow.

\section{Formalism for the inclusive rate}
The differential rate of semi-leptonic decays $B\to X_c\ell\bar\nu$
can be written as
$d\Gamma \propto |V_{cb}|^2 l^{\mu\nu}W_{\mu\nu}$
using the structure function (or hadronic tensor) $W_{\mu\nu}(p_B,q)$:
\begin{equation}
  W_{\mu\nu} = \sum_X (2\pi)^2\delta^{(4)}(p_B-q-p_X)
  \frac{1}{2M_B}
  \langle B(p_B)|J_\mu^\dagger(0)|X(p_X)\rangle
  \langle X(p_X)|J_\nu(0)|B(p_B)\rangle,
  \label{eq:structure_function}
\end{equation}
where the sum runs over all possible hadronic final states
$|X(p_X)\rangle$ with momentum $p_X$
\cite{Blok:1993va,Manohar:1993qn}.
The flavor-changing current $J_\mu=\bar{c}\gamma_\mu(1-\gamma_5)b$
induces the weak decay, and the momentum $q_\mu$ is transferred to the
lepton pair $\ell$ and $\bar\nu$.
The leptonic tensor $l^{\mu\nu}$ is a known kinematical factor.
Since the structure function is expressed as an imaginary part of the
matrix element
(sometimes called the Compton amplitude)
\begin{equation}
  T_{\mu\nu}(p_B,q) = \int\! d^4x e^{iqx}
  \frac{1}{2M_B}
  \langle B(p_B)|
  \mathrm{T}\left\{J_\mu^\dagger(0) J_\nu(0)\right\}
  |B(p_B)\rangle
  \label{eq:Compton_amplitude}
\end{equation}
of a product of currents
$\mathrm{T}\{J_\mu^\dagger(x)J_\nu(0)\}$, 
one can use OPE to approximate 
$\mathrm{T}\{J_\mu^\dagger(0) J_\nu(0)\}$
by a series of local operators, such as
$\bar{b}b$, $\bar{b} \bm{D}^2 b$, $\bar{b} \sigma\cdot\bm{B}b$, and so on.
It leads to an expansion in terms of inverse $b$ quark
mass $1/m_b$, since the momentum flowing into the charm quark
propagator is of order of $m_b$.
This forms a basis of the perturbative estimates of the
inclusive decay rate.
The matrix elements of the local operators sandwiched by the $B$ meson
states could be determined by fitting the experimental data of various
quantities. 
Some examples are shown, for instance, in \cite{Gambino:2013rza},
where the lepton energy and hadronic invariant mass moments are fitted
against the lepton energy cut applied in the experimental analysis.
Recent inclusive analysis includes the perturbative expansion to the
order of $\alpha_s^2$ and the heavy quark expansion to $1/m_b^3$
(see, for instance, \cite{Mannel:2019qel}).

Another important problem in the semi-leptonic $B$ meson decays is the
``missing component'' of the fully inclusive rate.
Namely, the sum of known exclusive decays such as
$B\to D^{(*)}\ell\bar\nu$, $D^{(*)}\pi\ell\bar\nu$, etc.
is less than the measurement of the total inclusive rate by about
15\%.
(See Table XIV of \cite{Dingfelder:2016twb}.)
This needs to be understood within the experimental analysis, but
theoretical information about the decay form factors to excited state
$D$ mesons would also be helpful.
Some discussions are found in the companion talk
\cite{Hashimoto_semileptonic}. 

Concerning the ``inclusive versus exclusive puzzle'',
there is also a theoretical question about the validity of the
quark-hadron duality.
Since the inclusive decays $B\to X_c\ell\bar\nu$ are dominated by the
$S$-wave states $D$ and $D^*$ to about 2/3 of the total decay rate,
there is not so much room left for the excited states.
The duality is expected to work only when the hadronic states are
sufficiently smeared, or summed, so that the details of the
bound-state dynamics become irrelevant.
Quantitative estimate of the duality violation effect is, however,
very difficult.

\section{Formalism for the lattice calculation}
On the lattice one can compute matrix elements of some operators.
For the calculation of semi-leptonic decay form factors of exclusive
processes, the matrix element of the form
$\langle D(p')|J_\mu|B(p)\rangle$
is computed.
Similarly the Compton amplitude (\ref{eq:Compton_amplitude}) can be
computed on the Euclidean lattice, but it corresponds to an 
unphysical kinematical setup.
For example, the hadronic energy $(p_B-q)_0$ accessible from the
Fourier transform of the Euclidean lattice data does not correspond to
the physical states; a strategy to connect the lattice data to some
physical observables was proposed in \cite{Hashimoto:2017wqo},
but the following method allows more direct calculation of
experimentally measured quantities.

An alternative approach is to view the Compton amplitude
(\ref{eq:Compton_amplitude}) as a spectral function.
One can write the integrand of the matrix element as
\begin{equation}
  \langle B(\bm{0})|\tilde{J}_\mu^\dagger(-\bm{q};t)
  \delta(\omega-\hat{H})
  \tilde{J}_\nu(\bm{q};0)|B(\bm{0})\rangle
  \label{eq:spectral}
\end{equation}
for a given energy $\omega=(p_B-q)_0$.
The currents are Fourier transformed to the momentum space;
the temporal direction remains in the coordinate space and
the energy $\omega$ is specified by the
$\delta(\omega-\hat{H})$ with the QCD Hamiltonian $\hat{H}$.
The inclusive rate can then be expressed as an integral of
(\ref{eq:spectral}) over $\omega$ and $\bm{q}^2$
with a weight factor (or a kernel function ) $K(\omega;\bm{q}^2)$ 
determined by the kinematical factor.
Once this type of the spectral function was extracted from the lattice
calculation, one can perform the integral over $\omega$ and
$\bm{q}^2$,
but the extraction of the spectral function from the Euclidean lattice
calculation is known as a ill-posed inverse problem and only some
approximate numerical solution can be obtained \cite{Hansen:2017mnd}.

A concrete proposal to bypass this problem is given in \cite{Gambino:2020crt}.
One realizes that the integral over the energy $\omega$ appearing in
the expression of the decay rate can be expressed as
\begin{equation}
  \int\! d\omega\,
  K(\omega;\bm{q}^2)
  \langle B(\bm{0})|\tilde{J}_\mu^\dagger(-\bm{q})
  \delta(\omega-\hat{H})
  \tilde{J}_\nu(\bm{q})|B(\bm{0})\rangle
  =
  \langle B(\bm{0})|\tilde{J}_\mu^\dagger(-\bm{q})
  K(\hat{H};\bm{q}^2)
  \tilde{J}_\nu(\bm{q})|B(\bm{0})\rangle.
  \label{eq:kernel_integral}
\end{equation}
On the other hand, the Euclidean correlation function calculated on
the lattice yields
\begin{equation}
  \langle B(\bm{0})|\tilde{J}_\mu^\dagger(-\bm{q})
  e^{-\hat{H}t}
  \tilde{J}_\nu(\bm{q})|B(\bm{0})\rangle.
  \label{eq:lattice_transfer}
\end{equation}
Therefore, if there is an approximation of the form
\begin{equation}
  K(\hat{H}) \stackrel{?}{=}
  k_0 + k_1 e^{-\hat{H}} + k_2 e^{-2\hat{H}} + \cdots + k_N  e^{-N\hat{H}},
  \label{eq:approx}
\end{equation}
one can relate the integrated decay rate and the lattice correlators.
Each term of the right hand side corresponds to the correlator of a
fixed time separation between the two current insertions.

The approximation of the form (\ref{eq:approx}) can be implemented
in various ways.
Essentially, the kernel $K(\omega;\bm{q}^2)$ can be considered as a
kind of smearing of the spectral function, as can be seen in
(\ref{eq:kernel_integral}), a weighted integral over $\omega$.
Solving the spectral function is an ill-posed problem, but once it is
smeared over some energy range, it may become much easier.
The form of the smearing cannot be controlled in the original form of
the Backus-Gilbert method \cite{Hansen:2017mnd},
but it was realized that the smearing can be specified by
including as a minimization in the Backus-Gilbert
method \cite{Hansen:2019idp}.
Another class of the approximation using an orthogonal polynomial
method was also proposed \cite{Bailas:2020qmv}.
We mainly discuss on this Chebyshev polynomial method in the
following. 

\section{Kernel approximation}
The kernel to be approximated has the following form:
\begin{equation}
  K(\omega) \sim e^{2\omega t_0} (m_B-\omega)^l
  \theta(m_B-|\bm{q}|-\omega).
  \label{eq:kernel}
\end{equation}
Here, the factor $e^{2\omega t_0}$ is introduced to keep a 
non-zero time separation $t_0$ between the two currents.
The configuration of having two currents in the equal Euclidean time
reflects the contributions from both $t>0$ and $t<0$, each of which corresponds to different
kinematical regions;
to obtain the semi-leptonic decays one has to restrict in $t>0$.
The next factor $(m_B-\omega)^l$ comes from the leptonic tensor and
$l$ = 0, 1 or 2.
The Heaviside function $\theta(m_B-|\bm{q}|-\omega)$ is introduced to
implement the kinematical upper limit of the $\omega$ integral.
The lower limit can be set to zero or to any value below the energy of
the lowest hadronic state.

\begin{figure}[tbp]
  \centering
  \includegraphics[width=7cm]{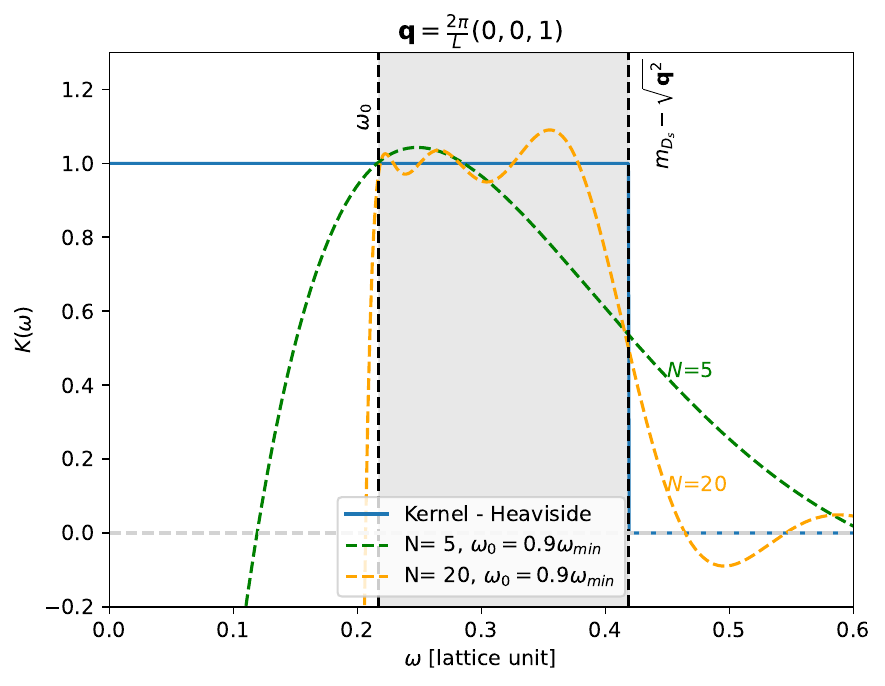}
  \includegraphics[width=7cm]{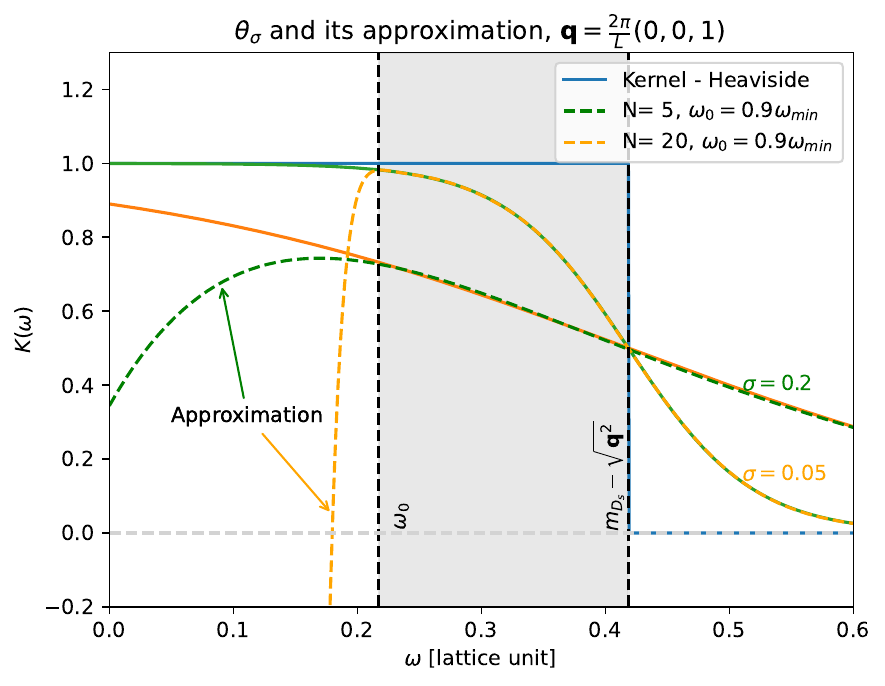}
  \caption{Chebyshev approximation of the Heaviside function
    $\theta(m_{D_s}-|\bm{q}|-\omega$.
    The order of the Chebyshev polynomial is set to $N=5$ (red) or
    $N=20$ (orange).
    The Heaviside function is directly approximated (left) or a
    sigmoid smoothing is introduced (right) with a smearing parameter
    $\sigma$ = 0.2 (red) or 0.05 (orange).
    The lower limit of the approximation is set to
    $\omega_0=0.9\omega_{\mathrm{min}}$, {\it i.e.} slightly below the
    kinematical lower limit.
    Plots from \cite{Kellermann:2022mms}.
  }
  \label{fig:kernel}
\end{figure}

An example of the kernel function and its approximation are shown in
Fig.~\ref{fig:kernel}. 
For clarity, we take the Heaviside function as a kernel and ignore the
factor $e^{2\omega t_0}$ and set $l=0$.
The kernel function has a discontinuity at the kinematical upper limit
$\omega=m_{D_s}-|\bm{q}|$ (Fig.~\ref{fig:kernel});
we introduce a smoothing (or smearing) to ease the polynomial
approximation by replacing the Heaviside function $\theta(x)$ by a sigmoid
function $\theta_\sigma(x)\equiv 1/(1+e^{-x/\sigma})$.
The parameter $\sigma$ represents the width of the smearing.
In the end, we have to take the limit $\sigma\to 0$.

The Chebyshev approximation is one of orthogonal polynomial
expansions.
(For the definitions and various physics applications, see
\cite{Weisse:2006zz}.)
In our case, it is written as
\begin{equation}
  \label{eq:chebyshev}
  K(\hat{H}) \simeq \sum_{j=0}^N c_j T_j(e^{-\hat{H}}),
\end{equation}
where $T_j(x)$ is the Chebyshev polynomial defined in the range
$[-1,+1]$.
(In practice, we use the shifted Chebyshev polynomial
$T_j^*(x)\equiv T_j(2x-1)$ so that $x$ is defined in $[0,1]$.)
The kernel operator is sandwiched by the $B$ meson state $|B\rangle$,
and it is evaluated with the operators $e^{-\hat{H}t}$ on the right hand
side appearing by expansing the Chebyshev polynomials.
Each term simply corresponds the (integer) time $t$ separation between
the two currents.

The expansion in (\ref{eq:chebyshev}) is mathematically well-defined
as $T_j(x)$ forms a orthogonal basis;
the coefficients $c_j$ can be easily calculated once the kernel
function $K(\omega)$ is given.
Moreover, the expansion is known to give the ``best'' approximation in
the sense that the maximum deviation in the range is minimal for a
given order $N$.

One important property of the Chebyshev polynomials is that the value
is confined in the region $|T_j(x)|\leq 1$.
It then leads to a constraint
$|\langle B|J^\dagger T_j(e^{-\hat{H}}) J|B\rangle/\langle B|B\rangle|\le 1$,
that comes from an obvious condition
$0\leq |e^{-\hat{H}}|\leq 1$ for the Hamiltonian eigenvalues.
Combining with the expansion (\ref{eq:chebyshev}), this constraint
gives a strict upper- and lower-limits of the estimate by taking
$\langle B|J^\dagger T_j(e^{-\hat{H}}) J|B\rangle/\langle B|B\rangle=\pm 1$
as two extreme cases for the ignored terms $j>N$.
The systematic error due to the approximation can thus be estimated
strictly as $\sum_{j=N}^\infty |c_j|$.
(Remember that the $c_j$'s are calculated at low cost, and $c_j$ typically
dumps exponentially for large $j$.)
This is an advantage over other methods such as that of
\cite{Hansen:2019idp}.

\section{Early calculations}

\begin{figure}[tbp]
  \centering
  \includegraphics[width=9cm]{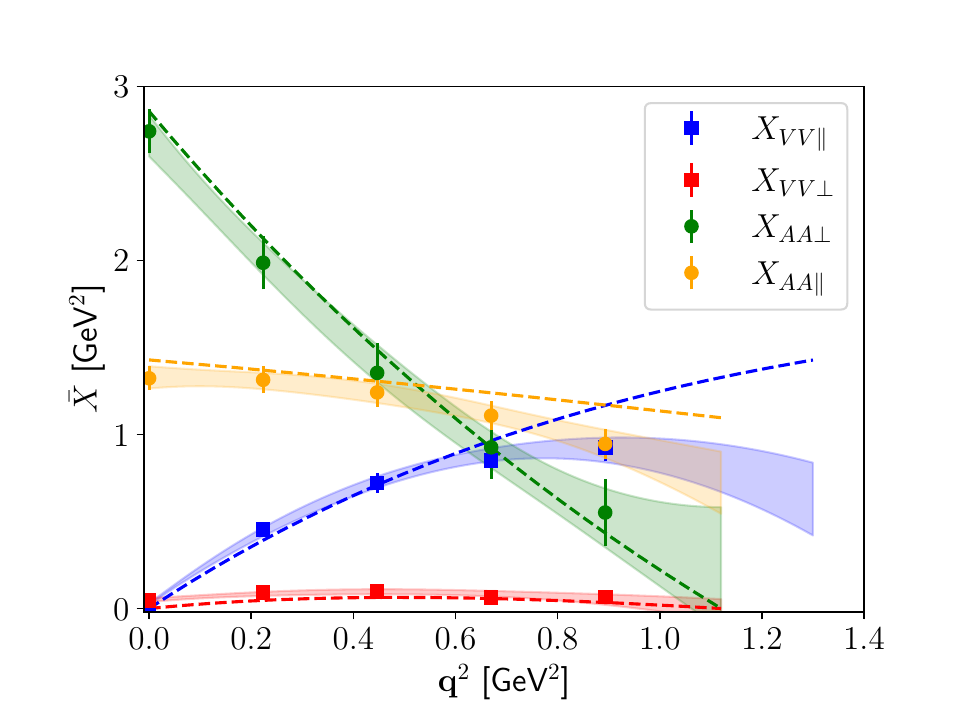}
  \caption{
    $B_s\to X\ell\bar\nu$ differential decay rate (divided by $|\bm{q}|$).
    An example from \cite{Gambino:2022dvu} computed with $m_b$ lighter
    than the physical value.
    Different colors show the decomposition into four different
    channels: vector ($VV$) or axial-vector ($AA$) current;
    parallel ($\parallel$) or perpendicular ($\perp$) polarization.
  }
  \label{fig:diff_decayrate}
\end{figure}

An early result for the differential decay rate (divided by
$|\bm{q}|$) from \cite{Gambino:2020crt,Gambino:2022dvu} 
is shown in (\ref{fig:diff_decayrate}).
Here the initial $b$ quark is lighter than its physical value
(about 3~GeV), so that
the kinematical upper-limit for $\omega$ is also made lower.
The results for different channels are plotted separately as a
function of the recoil momentum squared $\bm{q}^2$.
The channels are those of vector or axial-vector current insertions,
and of different orientation of the currents (parallel or
perpendicular to $\bm{q}$).

Also plotted by the dashed line is the expectation from the ground
state (the $S$-wave $D$ and $D^*$ mesons) contributions.
They are obtained from the form factor calculations associated with
the work \cite{Colquhoun:2022atw,Aoki:2023qpa}.
It turned out that the inclusive rate is saturated by the ground
states, as anticipated due to the smaller phase-space compared to the
physical setup because of the smaller $b$-quark mass.
By inspecting the individual matrix elements
$\langle B|J^\dagger e^{-\hat{H}t}J |B\rangle/\langle B|B\rangle$,
we can identify the excited-state contributions, but their size is
insignificant at the scale of this plot.

The results are compared with OPE in \cite{Gambino:2022dvu}, where the
perturbative calculation is performed to $O(\alpha_s)$ and the power
corrections are included to $O(1/m_b^3)$.
They are in good agreement, although the systematic error of the OPE
calculation is large because the lattice calculation is done at smaller
$b$ quark mass, where the power corrections are enhanced.

\begin{figure}[tbp]
  \centering
  \includegraphics[width=9cm]{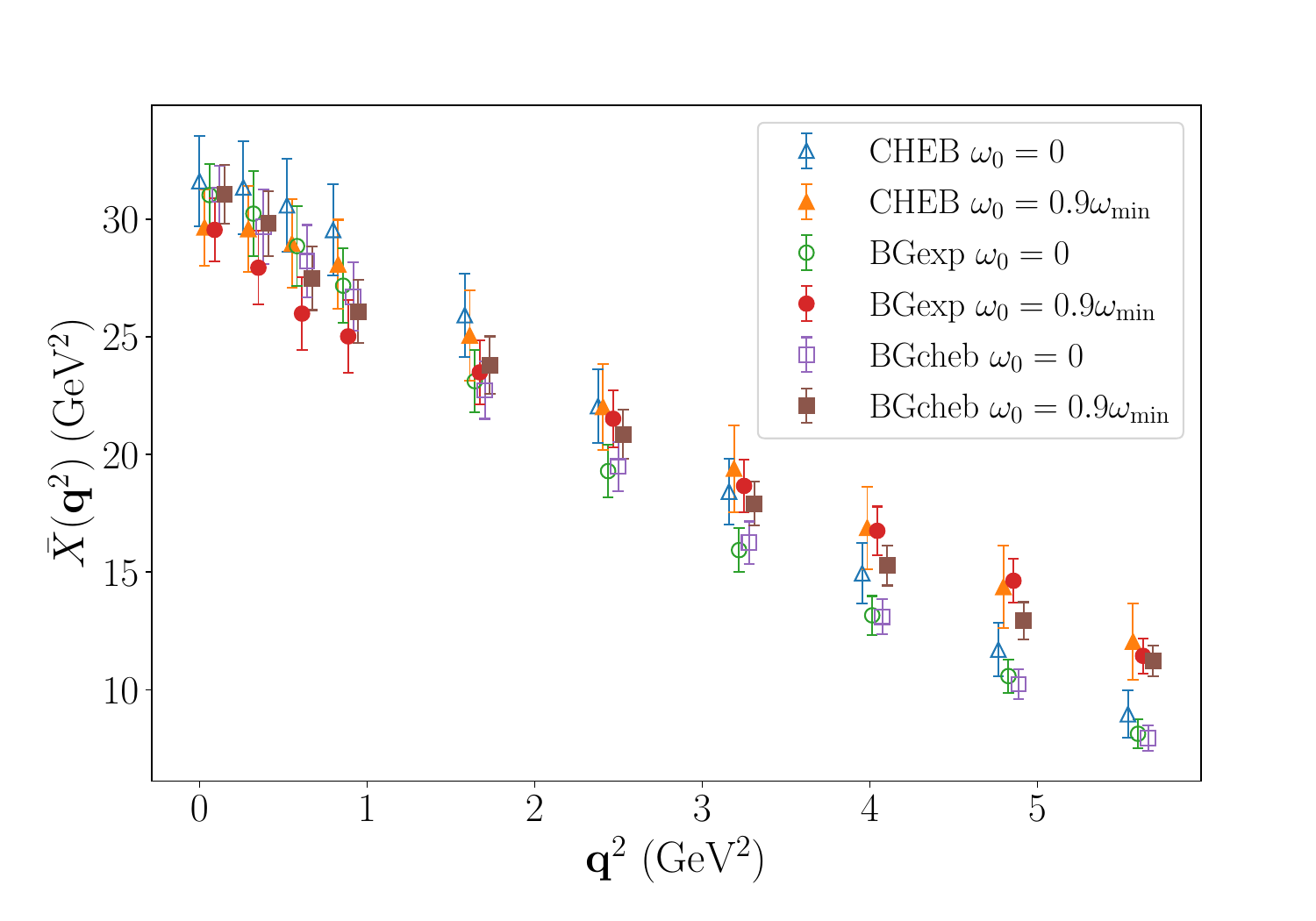}
  \caption{
    $B_s\to X\ell\bar\nu$ differential decay rate (divided by
    $|\bm{q}|$) at the physical $b$ quark mass.
    The data are from \cite{Barone:2023tbl}.
    All channels are combined.
    Kernel approximation is performed with the Chebyshev polynomial
    (CHEB), the Backus-Gilbert method (BGexp), or the Backus-Gilbert
    method in the Chebyshev basis (BGcheb).
  }
  \label{fig:Xbar}
\end{figure}

First calculation at the physical $b$ quark mass is performed by
\cite{Barone:2023tbl}.
In Fig.~\ref{fig:Xbar} the differential decay rate (divided by
$|\bm{q}|$) is plotted for different choices of the kernel
approximations, basically the Chebyshev approximation and
Backus-Gilbert method.
The results are in good agreement among the methods.
Indeed, we confirmed that the resulting approximation curve of the
kernel function is very similar between Chebyshev and Backus-Gilbert.
A slight deviation among different setup is visible at large values of
$\bm{q}^2$ depending on the choice of $\omega_0$, the lower limit of
the approximation. 
It suggests that there is still some hidden uncertainty in the kernel
approximation.

\section{Systematic errors}
This brings us to more detailed study of the systematic errors.
In particular, the largest momentum point of $X_{VV\parallel}$ in
Fig.~\ref{fig:diff_decayrate} (the blue point around $\bm{q}^2$ =
0.9~GeV$^2$) may suggest a problem:
the inclusive rate is significantly lower than the ground state ($D$
meson) contribution.

The same channel, but for the $D_s$ meson decays, is studied in 
\cite{Kellermann:2022mms}.
Near the maximum $\bm{q}^2$ the kernel function has a narrow
support from the lowest energy state $\sqrt{m^2+\bm{q}^2}$ up to the
upper limit $m_{D_s}-|\bm{q}|$.
(The kinematical endpoint is defined by the limit where the lowest and
the upper limit meet.)
As one can imagine from Fig.~\ref{fig:kernel} the kernel approximation gets
harder in this limit.
Roughly speaking, the polynomial of order $N$ allows the slope of the
approximated kernel of about $O(N)$, which corresponds to the
Heaviside-like change within the range of $\omega\sim 1/N$.
When the support of the kernel becomes as small as this width, the
approximation would become imprecise.

\begin{figure}[tbp]
  \centering
  \includegraphics[width=9cm]{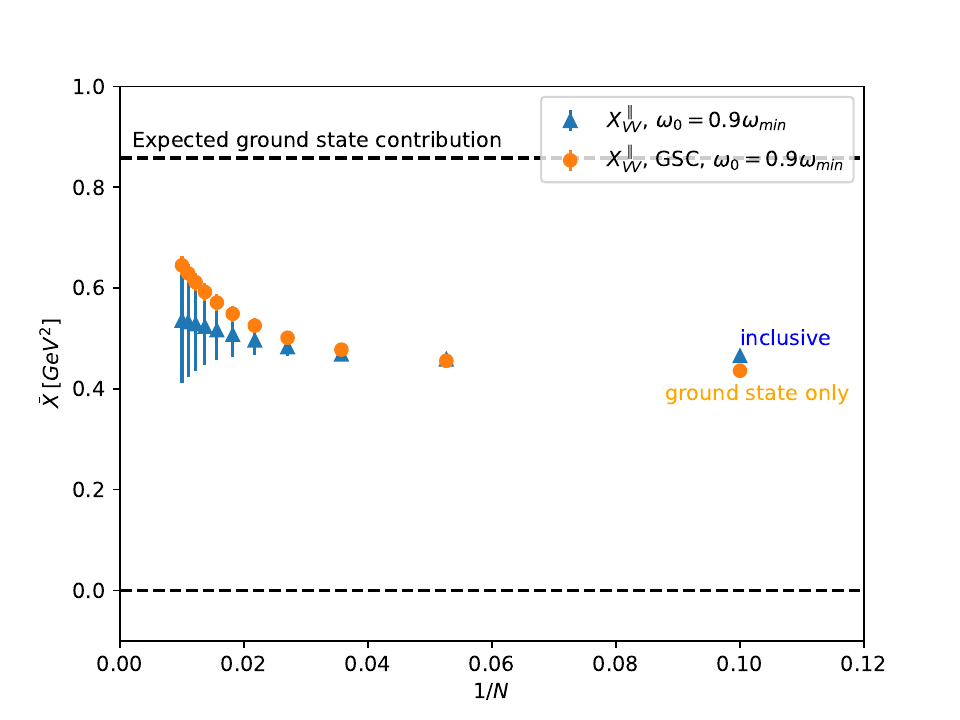}
  \caption{
    Differential decay rate near the kinematical end-point for the
    channel of $VV$ insertion parallel to $\bm{q}$.
    The plot is against the smearing width $\sigma=1/N$.
    The inclusive calculation (triangle) is plotted with a bound
    obtained by setting the Chebyshev matrix elements of $j$ larger
    than 10 to $\pm 1$.
    The corresponding ground-state contribution is given by circles,
    and its limit to $\sigma\to 0$ is shown by a dashed line.
  }
  \label{fig:XVV}
\end{figure}

In order to control such systematic effect, we introduce a 
smeared Heaviside function with a width $\sigma$ and set it to
$\sigma=1/N$.
Then, the Chebyshev approximation at the order $N$ follow the smeared 
kernel rather precisely.
Even though the lattice data are not available for large time
separation, thus large $N$, one can obtain the upper and lower limit
of the estimate for arbitrary $N$ by setting their Chebyshev matrix
elements of unknown higher orders to $\pm 1$ as mentioned above.
It is done in Fig.~\ref{fig:XVV}, and the estimate is shown as a
function of $\sigma=1/N$.
Towards the limit of $\sigma\to 0$, one can see that the error
increases for the (differential) inclusive rate.
Also plotted is its contribution from the ground state, for which the
energy is precisely known and the distortion due to the kernel
approximation can be traced.
We find that the inclusive rate actually covers the ground-state
contribution for each value of $\sigma=1/N$; the limit of
$\sigma\to 0$ may be estimated by an extrapolation.
One may think that the error in the $\sigma\to 0$ limit is too large
to be useful, but one can avoid the problem for this particular case
because we know that the ground state saturates the decay rate.

Another potentially important source of the systematic error is the
finite volume effect.
The excited states consist of multi-hadron final states, and their
spectrum becomes discrete in a finite box and a systematic error of
$1/L$ is expected.
The size of such error can be estimated by assuming the two-body
spectrum and their amplitude \cite{Kellermann:2023yec}.

\section{Final remarks: towards understanding the puzzle}
As outline above, there is a formulation to compute the inclusive
decay rate using lattice QCD.
It doesn't require the reconstruction of the spectral function, and
one can avoid the ill-posed inverse problem.
The new method involves new problems, though.
The kernel approximation is challenging when the hard energy cutoff
has to be implemented like the case of the inclusive decay rate.
A good news is that the systematic error can be rigorously estimated
with the Chebyshev polynomial method.

Some initial lattice studies are encouraging.
Comparison with the OPE-based approach has been made; the study
should be extended to the calculations with the physical kinematics.
Such consistency check would finally establish the consistency among
the theories of inclusive decays, {\it i.e.} OPE and lattice.
Eventually, the comparison should also be made for various moments,
such as those of lepton energy and hadronic mass.
Lattice calculation of them can be performed in parallel with the
decay rate.

More detailed comparison among the OPE, lattice and experiments are
possible. 
In addition to the total decay rate and moments, one can consider 
any weighted integrals of the differential decay rate.
Each party (OPE, lattice, experiment) has its own advantages and
weaknesses, and one can reach the optimal solution after compromises.
For instance, the lattice calculation is harder and thus less precise
for large recoil momenta, while the experiment is not able to cover
too small momenta.
The OPE method needs a sufficiently broad range of integral to avoid
potential problems due to the quark-hadron duality.
Close collaborations among different parties would be crucial to
finally resolve the puzzle.

\vspace*{5mm}
I thank the present and past members of the JLQCD collaboration,
as well as other collaborators that led to the publications
\cite{Gambino:2022dvu,Barone:2023tbl}.
A lot of materials in this presentation emerged from the discussions
with them.

This work is partly supported by MEXT as ``Program for Promoting
Researches on the Supercomputer Fugaku'' (JPMXP1020200105)
and by JSPS KAKENHI, Grant-Number 22H00138.
This work used computational resources of supercomputer
Fugaku provided by the RIKEN Center for Computational Science
through the HPCI System Research Projects
(Project IDs: hp120281, hp230245),
SX-Aurora TSUBASA at the High Energy Accelerator Research
Organization (KEK) under its Particle, Nuclear and Astrophysics
Simulation Program
(Project IDs: 2019L003, 2020-006, 2021-007, 2022-006 and 2023-004).

\end{document}